# AN ITERATIVE METHOD APPLIED TO CORRECT THE ACTUAL COMPRESSOR PERFORMANCE TO THE EQUIVALENT PERFORMANCE UNDER THE SPECIFIED REFERENCE CONDITIONS

Yuanyuan Ma, Harald Fretheim, Erik Persson, and Trond Haugen

Department of Product and Technology for Oil, Gas, and Petrochemicals,
Division of Process Automation, ABB, Oslo, Norway
{yuanyuan.ma;harald.fretheim;erik.persson;trond.haugen}@no.abb.com

## ABSTRACT

*This paper proposes a correction method, which corrects the actual compressor performance in real operating conditions to the equivalent performance under specified reference condition. The purpose is to make fair comparisons between actual performance against design performance or reference maps under the same operating conditions. Then the abnormal operating conditions or early failure indications can be identified through condition monitoring, which helps to avoid mandatory shutdown and reduces maintenance costs. The corrections are based on an iterative scheme, which simultaneously correct the main performance parameters known as the polytropic head, the gas power, and the polytropic efficiency. The excellent performance of the method is demonstrated by performing the corrections over real industrial measurements.*

## KEYWORDS

*Corrections, Actual Operating Conditions, Reference Operating Conditions, Reference Map, Compressor Performance, Corrected Performance, Expected Performance*

## 1. INTRODUCTION

Compressor performance is dependent on many external process conditions such as pressures, temperatures, and gas compositions [1, 2]. Monitoring the compressor performance can identify abnormal compressor performance and problems at an early stage so that failures or mandatory shut down can be avoided, which significantly reduces maintenance cost and is essential for daily operation actions [3].

One main challenge in condition monitoring is how to fully utilize the monitored performance parameters in further analysis and troubleshooting [4]. When comparing the actual performance against design/reference performance, the comparisons must be fairly and taken under the same operating conditions. The design performance or reference maps are often conducted under certain specific inlet pressure, temperature, and gas composition, which are known as reference conditions. In contract, the actual performance is only valid for their real operating conditions, which varies for each operating point. To compare these performances under the same operating conditions, one solution is to transform the actual performance in real conditions to the equivalent performance valid for the reference conditions. This task is known as corrections and is the main focus of the paper.

To the best of the authors' knowledge, such corrections have been presented so far only in [5] for validating a new centrifugal compressor's compliance to the guarantee conditions after it is built. To check whether the compressor will meet the specified duty, the performance achieved from a shop test needs to be compared with specifications provided by a manufacturer.

However, due to practical difficulties, the test gas conditions are usually different with the specific gas conditions used previously by a manufacturer. To compare the performance equivalently, the rules for correlating the results obtained with the test gas to that with the design specifications were proposed in [5].

Although three different testing cases have been discussed in [5], they are only categorized by various testing gas conditions. The reference conditions can be characterized not only by the gas conditions but also by pressures and temperatures. Unfortunately, the correction method in [5] does not work for a more general case and definitely limits the applications of the method. Moreover, the existing correction method requires a big set of parameters and meter readings to complete the whole correction process. In reality, rather than in a test, full access of all these information is not practical. Such a method cannot fulfil basic applications in condition monitoring such as presenting actual compressor performance in a reference map.

To eliminate all limitations mentioned above and extend the applications of the corrections, we propose a new and practical correction method for real gas. The method can correct actual compressor performance to more practical reference conditions, which are specified not only by the gas composition, but also by the inlet pressure and temperature. The method is based on an iterative scheme, which simultaneously corrects the main performance parameters including the polytropic head, the gas power, and the polytropic efficiency. The report `ASME PTC-10' [5] only presents the instructions how to carry out the corrections. In our paper, we also validate the proposed method by performing the corrections over the real operating points sampled from our industrial gas processing pump pilot. It will be shown experimentally that the corrected performance are very close to the expected values obtained directly from a given reference map. It is also interesting to mention that since our new method requires less information to carry out the corrections, it is very easy to compare two sets of historical data by correcting them to pre-defined reference conditions.

The paper is organized as follows. Section 2 describes the general procedure of the corrections. The principle regarding how to correct an individual operating point under the real conditions to the specific reference conditions is introduced in Section 3. The verification of the correction methods is discussed in Section 4. Results for demonstrating the accuracy of the proposed method are presented in Section 5. Finally, some concluding remarks are given in Section 6.

## 2. GENERAL PROCEDURE OF CORRECTIONS

Section 2 describes the general procedure and the basic assumptions that applied to the corrections.

Figure 1 explains the idea regarding how to correct an individual operating point under the real conditions to the specific reference conditions:

1) The first step is to transform the real operating conditions, i.e., the inlet & discharge pressures, temperatures, and gas composition, to the specific reference conditions. As shown in Fig. 1, the inlet pressure and temperature of the operating point after corrections are considered to be the same as the reference inlet ones. The gas composition for the operating point is also changed from the real one to the reference gas composition.

2) The second step of the correction is to find out, under certain constraints, what the equivalent discharge pressure and temperature are when the real inlet conditions change to the reference conditions.

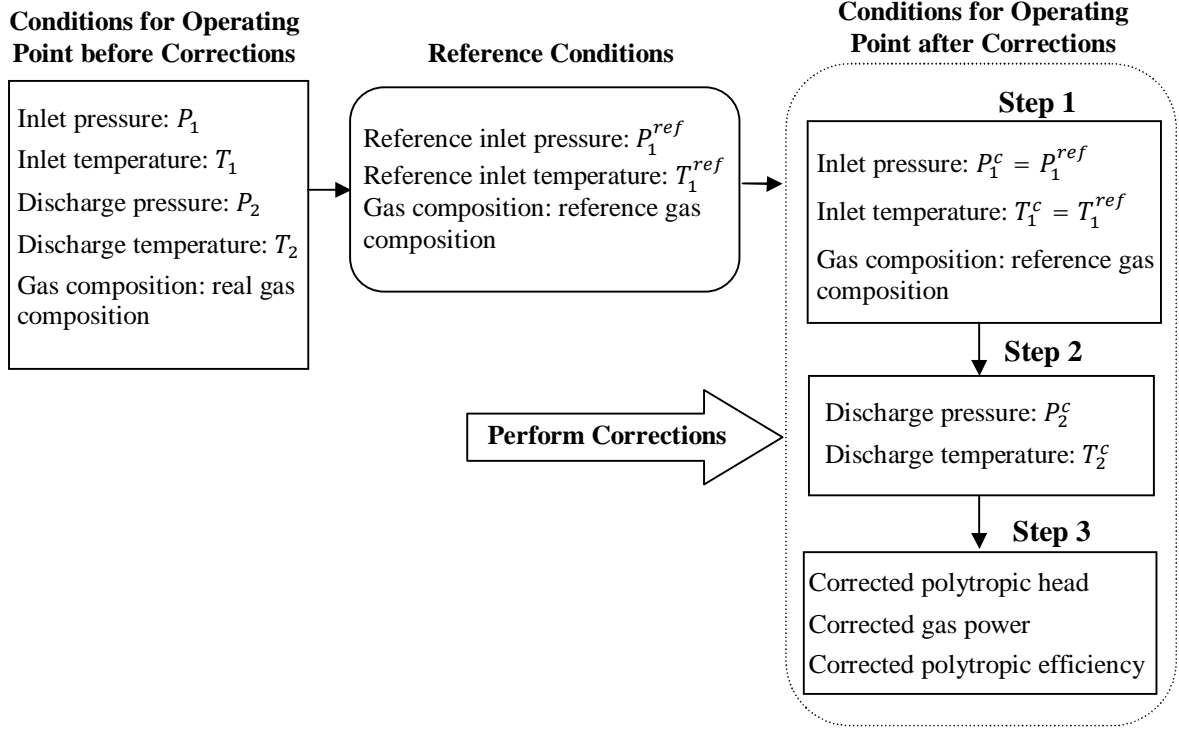

Figure 1 Interpretation of Correction Principle.

3) Finally, recalculate the polytropic head, gas power, and polytropic efficiency based on the corrected gas composition as well as the corrected inlet & discharge pressures and temperatures. The new values calculated under the reference conditions are referred to corrected performance.

The constraints mentioned in Step 2 are known as follows:

1) The polytropic efficiency is constant before and after the corrections;

2) The ratio of the volumetric flow at inlet to the flow at outlet remains the same before and after the corrections.

The first assumption comes from the fact that the polytropic efficiency is independent on the thermodynamic state such as the temperature and the pressure [6]. The purpose of the second assumption is to have similar flow conditions before and after the corrections. It indicates that the density ratio before and after the corrections remains the same.

## 3. CORRECTION METHODOLOGY

As discussed in Section 2, the corrected inlet conditions of operating points are assumed to be equal to the given reference conditions. The remaining problem is how to determine the corrected discharge conditions based on the known pressures, temperatures, and gas compositions. This is the most challenge part in the corrections.

The procedure to determine the corrected discharge values and the corrected performance can be summarized as follows:

1) Calculate the average physical gas properties of the original operating points including: average polytropic exponent $n$, average specific heat ratio $k$, average additional function $X$ and $Y$, which are given by

$$n = \frac{n_1 + n_2}{2}, \tag{1}$$

$$k = \frac{k_1 + k_2}{2}, \tag{2}$$

$$X = \frac{X_1 + X_2}{2}, \tag{3}$$

$$Y = \frac{Y_1 + Y_2}{2}. \tag{4}$$

Here, the symbols $n_1, k_1, X_1$, and $Y_1$ denote the polytropic exponent, specific heat ratio, and additional functions at inlet, while the four parameters at discharge are represented by $n_2, k_2, X_2$, and $Y_2$. The two additional functions mentioned here are used to supplement the compressibility factor calculated under real-gas conditions [7].

These parameters can be calculated by the function *'ExponentCalc'* introduced in Appendix A for real gas. The required inputs for using this function are the relevant pressure, temperature, and gas composition.

2) Calculate the corrected inlet polytropic exponent and the specific heat ratio.

Since the inlet pressure and temperature have been changed to the reference ones after the corrections, the physical gas properties at inlet, known as the corrected inlet polytropic exponent $n_1^c$, corrected specific heat ratio $k_1^c$, and two corrected additional functions $X_1^c$, and $Y_1^c$, need to be recalculated. The procedure and equations can be found in Appendix A.

3) Calculate the corrected discharge pressure, temperature, and average polytropic exponent.

The first assumption indicates that the corrected polytropic efficiency ($\eta^c$) equals to the uncorrected polytropic efficiency ($\eta$), i.e.,

$$\eta^c = \eta. \tag{5}$$

According to the definition of the polytropic exponent for real-gas conditions [7], the polytropic efficiency can be written as:

$$\eta = \frac{Y \cdot n \cdot (1-k)}{k \cdot (1+X) - Y \cdot n \cdot (k+X)}. \tag{6}$$

Substituting (6) and the similar expression for the corrected polytropic efficiency in eq. (5) yields the relationship (7), which is presented at the bottom of this page.

$$\frac{Y \cdot n \cdot (1-k)}{k \cdot (1+X) - Y \cdot n \cdot (k+X)} = \frac{Y^c \cdot n^c \cdot (1-k^c)}{k^c \cdot (1+X^c) - Y^c \cdot n^c \cdot (k^c + X^c)}. \tag{7}$$

The symbols $X^c$, $Y^c$, and $k^c$, in (7) represent the two average corrected additional functions and the average corrected specific heat ratio. The corrected polytropic exponent $n^c$ can then be deduced from (7) and expressed in the form as shown in (8).

$$n^c = \frac{Y \cdot n \cdot (1-k) \cdot k^c \cdot (1+X^c)}{[k \cdot (1+X) - Y \cdot n \cdot (k+X)] \cdot Y^c \cdot (1-k^c) + Y \cdot n \cdot (1-k) \cdot Y^c \cdot (k^c + X^c)}. \quad (8)$$

Since these average corrected values $X^c$, $Y^c$, and $k^c$, on the right side of equation (7) are unknown and their calculations depend on the value for $n^c$, equation (8) cannot be solved.

Our solution is to replace these average corrected parameters $(X^c, Y^c, k^c)$, in (8) by the parameters $(X_1^c, Y_1^c, k_1^c)$ at inlet. Then we use the new obtained value, denoted by $n_{initial}^c$ (see (9)), as the initial value and calculate the corrected polytropic exponent $n^c$ through the following iterative scheme:

$$n_{initial}^c = \frac{Y \cdot n \cdot (1-k) \cdot k_1^c \cdot (1+X_1^c)}{[k \cdot (1+X) - Y \cdot n \cdot (k+X)] \cdot Y_1^c \cdot (1-k_1^c) + Y \cdot n \cdot (1-k) \cdot Y_1^c \cdot (k_1^c + X_1^c)}. \quad (9)$$

**Step 0:** Set the initial value $n_{initial}^c$ as the corrected polytropic exponent, i.e.,

$$n^c = n_{initial}^c. \quad (10)$$

**Step 1:** Calculate the corrected discharge pressure and temperature.

According to Appendix B, the corrected pressure and temperature are given by

$$P_2^c = P_1^c \cdot \left(\frac{P_2}{P_1}\right)^{\frac{n^c}{n}}, \quad (11)$$

$$T_2^c = T_1^c \cdot \left(\frac{T_2}{T_1}\right)^{\frac{n^c-1}{n-1}}. \quad (12)$$

**Step 2:** Calculate the corrected discharge compressibility factor $z_2^c$, enthalpy $h_2^c$, specific heat capacity $c_{p2}^c$ and molecular weight $MW^c$ for the corrected operating points. The physical properties can be computed according to the expressions presented in [8].

**Step 3:** Calculate the corrected polytropic exponent at discharge, which is expressed as

$$n_2^c = \frac{1 + X_2^c}{Y_2^c \cdot \left[\frac{1}{k_2^c} \cdot \left(\frac{1}{\eta^c} + X_2^c\right) - \left(\frac{1}{\eta^c} - 1\right)\right]}. \quad (13)$$

Here, $X_2^c$, $Y_2^c$, and $k_2^c$ can be computed according to the function *'ExponentCalc $(P_2^c, T_2^c, c_{p2}^c, z_2^c, \eta^c$, reference gas composition)'* presented in Appendix B.

**Step 4:** Calculate the overall corrected polytropic exponent by averaging the corrected polytropic exponents at inlet and discharge, i.e.,

$$n^c = \frac{n_1^c + n_2^c}{2}. \tag{14}$$

**Step 5:** Repeat Steps 1-4 for 100 times. The values obtained after the iterations for $P_2^c$, $T_2^c$, and $n^c$ are considered as the corrected discharge pressure, temperature, and polytropic exponent.

4) Calculate the Schultz factor f by the equation below [5,7]

$$f = \frac{(k_s^c - 1) \cdot (h_{2s}^c - h_1^c)}{k_s^c \cdot (z_{2s}^c \cdot R \cdot T_{2s}^c / MW_s^c - z_1^c \cdot R \cdot T_1^c / MW^c)}, \tag{15}$$

where

$$k_s^c = \frac{1}{2} \cdot \left( \frac{k_1^c}{Y_1^c} + \frac{k_2^c}{Y_2^c} \right), \tag{16}$$

denotes the corrected isentropic exponent. In (15), $R = 8314.3 Pa\ m^3/kmol/K$, $h_1^c$, and $z_1^c$ are the gas constant, corrected enthalpy, and corrected compressibility at inlet. The values for $h_1^c$, and $z_1^c$ can be calculated by the theory introduced in [8]. The symbols with subscript 's' denote the parameters of isentropic process.

5) Calculate the corrected polytropic head $H_p^c$ [9]:

$$H_p^c = f \cdot \frac{n^c}{n^c - 1} \cdot \frac{1}{MW^c} \cdot (z_2^c \cdot R \cdot T_2^c - z_1^c \cdot R \cdot T_1^c). \tag{17}$$

6) Correct the speed and the mass flow according to the fan law [10, 12]:

$$N^c = N \cdot \sqrt{\frac{H_p^c}{H_p}}, \tag{18}$$

$$m^c = m \cdot \frac{N_c}{N}. \tag{19}$$

In (18), $N^c$, $N$ and $Hp$ are known as the corrected compressor speed, compressor speed, and polytropic head. The calculations for the polytropic head can be found in [6]. In (19), the symbols $m^c$ and $m$ are corrected mass flow and original mass flow.

7) Calculate the corrected gas power [6]

$$PWR^c = m^c \cdot \frac{H_p^c}{\eta^c}. \tag{20}$$

## 4. VERIFICATION OF THE CORRECTION METHODOLOGY

The accuracy of the correction method can be verified by the relative difference between the corrected performance and the expected performance. The interest performance parameters for verifications are polytropic head and gas power. The expected performance are the output values predicted directly from the given compressor performance maps. If the performance has been corrected to the reference conditions properly, the corrected performance is supposed to be close to the expected one.

The deviation of the corrected polytropic head or gas power from the expected one are defined by

$$\delta_{H_p} = \left|\frac{H_p^c - H_p^e}{H_p^c}\right| \times 100\%, \tag{21}$$

$$\delta_{PWR} = \left|\frac{PWR^c - PWR^e}{PWR^c}\right| \times 100\%. \tag{22}$$

In the equations above, the symbol $H_p^e$ and $PWR^e$ denote the expected polytropic head and gas power. The expected performance can be determined by locating the expected operating point from a given reference map. The location of the point can be found by first mapping the given mass flow to the performance characteristic curved curve at specified compressor speed. Then, the expected value can be determined by mapping the expected operating point to the y-axis describing the performance parameter, e.g., the corrected polytropic head.

## 5. RESULTS

As To verify the correctness of the proposed correction method, the deviation of the corrected performance from the expected performance predicted in a reference map will be studied. The measurements used in verifications are real industry gas processing pump data, which record the compressor actions for the past three years.

If a reference map is already given, one just needs to follow the procedure introduced in Section 3 to carry out the corrections for all real operating points. Then, determine the expected performance as discussed in Section 4. Otherwise, if the reference map is not available, which is the case for this paper, a reference map needs to be generated from certain amount of the real operating points. First correct certain operating points to some pre-set reference conditions. Then apply the third polynomial curve fitting technique to generate the reference map.

In our case, the operating points sampled at 2011 were used to generate the reference map. The reference inlet pressure and temperature were set to $P_1^{ref} = 76.5$ bar and $T_1^{ref} = 299.5 K$.

Figures 2-4 illustrate the deviations between the corrected polytropic head and the expected one for different operating period.

Figure 2 shows the correction results performed on the operating points from 2011. These operating points are the same data as the one used to generate the reference map. Due to this fact, the obtained deviations are supposed to approach to zero. This is illustrated by the results presented in Fig. 2. As can be seen, the deviations are very small and the average deviation is less than 1%.

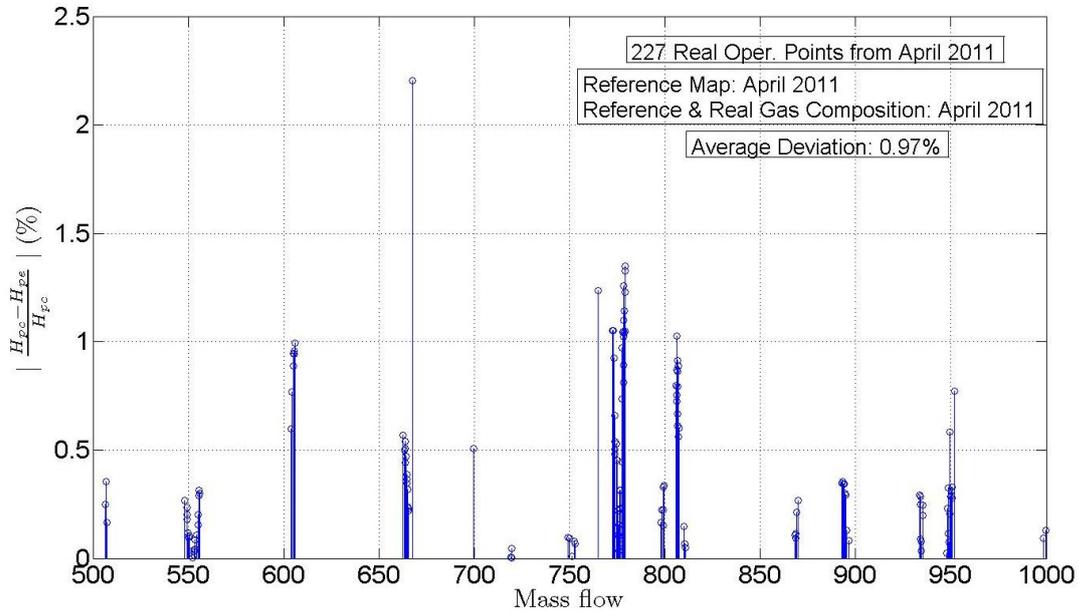

Figure 2 Corrected Polytropic Head Deviations for Operating Data from 2011.

Figures 3 and 4 present the correction results for operating points sampled at 2009 and 2012. Even the operating conditions especially the real gas composition is different with the reference one, the correction method still works and returns acceptable results. The average deviation is 3.67% when correcting the operating data at 2009 and the average deviation is 3.53% when the correcting data at 2012.

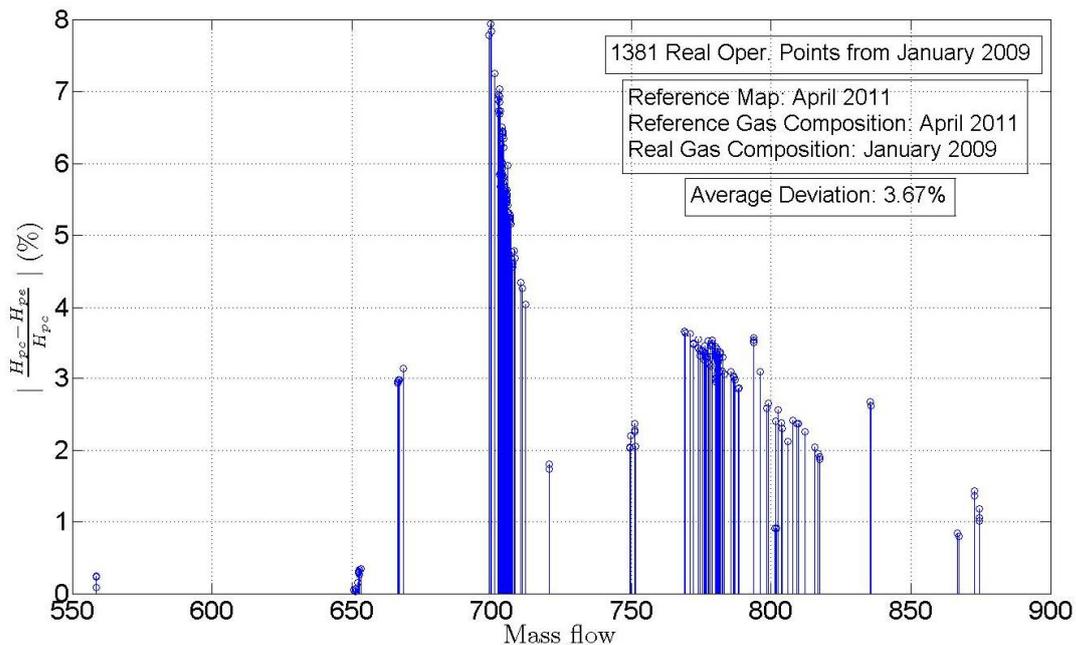

Figure 3 Corrected Polytropic Head Deviations for Operating Data from 2009.

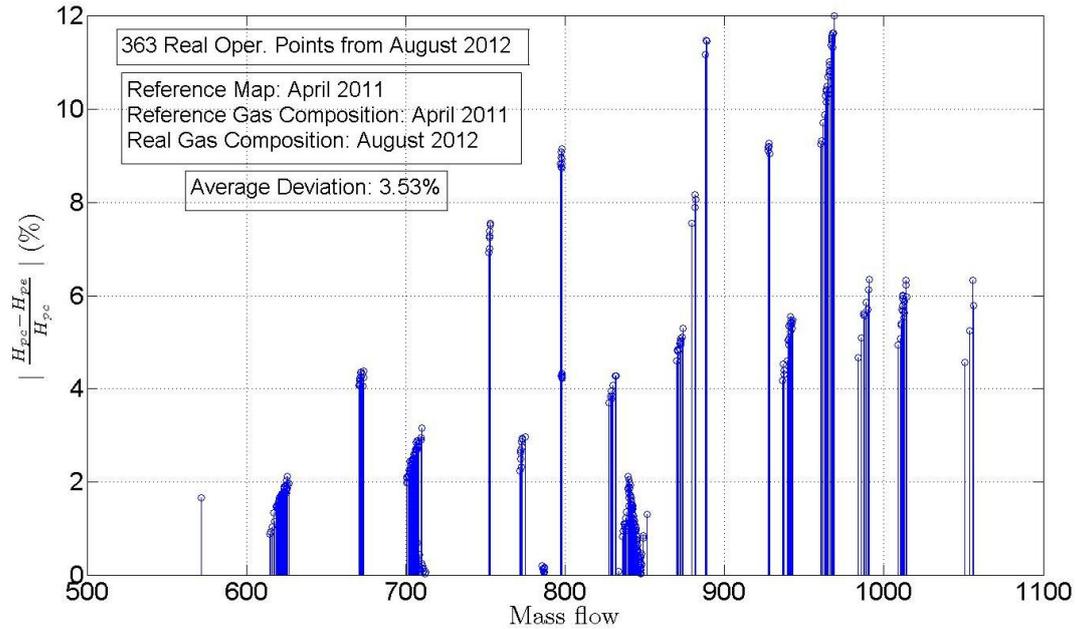

Figure 4 Corrected Polytropic Head Deviations for Operating Data from 2012.

Some peak values exist in Figs. 3 and 4, which indicate that large deviations exist in the corrections. There are two reasons for getting such peak values. One reason is that for some data points, their operating conditions are quite different with the reference conditions. It is already beyond the correction capability. Then big deviations occur. The conditions here can be the inlet (discharge) temperatures, pressures, and gas composition. For example, in our further study, we find that when the inlet pressure or the discharge pressure is 10% away from the reference pressure, the average deviation exceeds 10%. The other factor causing large deviations comes from the accuracy and the valid range of the reference map. In our paper, the reference map is not a standard one provided by manufactures. It is generated by applying the polynomial fitting techniques on a certain amount of the operating points. Then, the accuracy of the fitting techniques directly influences the accuracy of the map. Moreover, since the reference map is generated based on a limit number of the operating points. The map has its own valid range. If the operating points after the corrections are already out of the valid range, the expected performance obtained from the map is not accurate. This leads to severe deviations from the corrected values. Here, since they are beyond the scope of the paper, we will not discuss more in the paper. In general, the correction method is efficient in correcting operating points with all kinds of conditions. Even counting these peak values and the factors which bring inaccuracy in evaluating the correction method, the average deviation is still low.

Figures 5-7 show the corrected gas power deviations for the operating points sampled at 2009, 2011, and 2012. The average deviations are 2.08%, 0.68%, 2.6%, respectively. The average deviations for the corrected gas power are even smaller than the deviations for the corrected polytropic head. More important, all figures presented above verify that the correction method performs well when simultaneously correcting the two main performance parameters: the polytropic head and the gas power. In addition, in corrections, we assume the polytropic efficiency remains the same before and after corrections, the deviations for corrected efficiency are always zero.

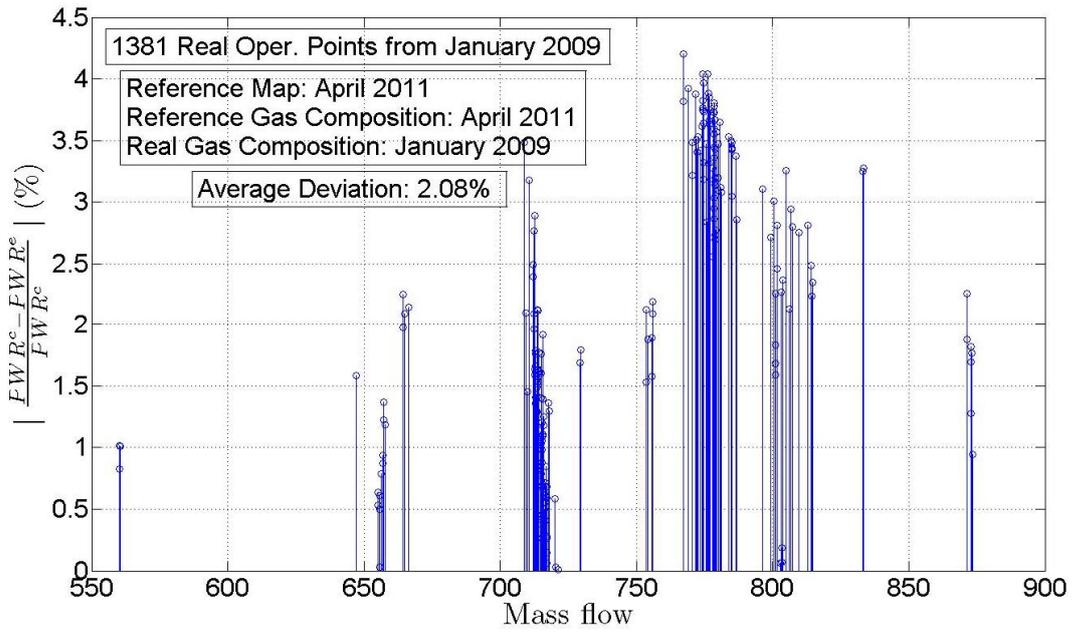

Figure 5 Corrected Gas Power for Operating Data from 2009.

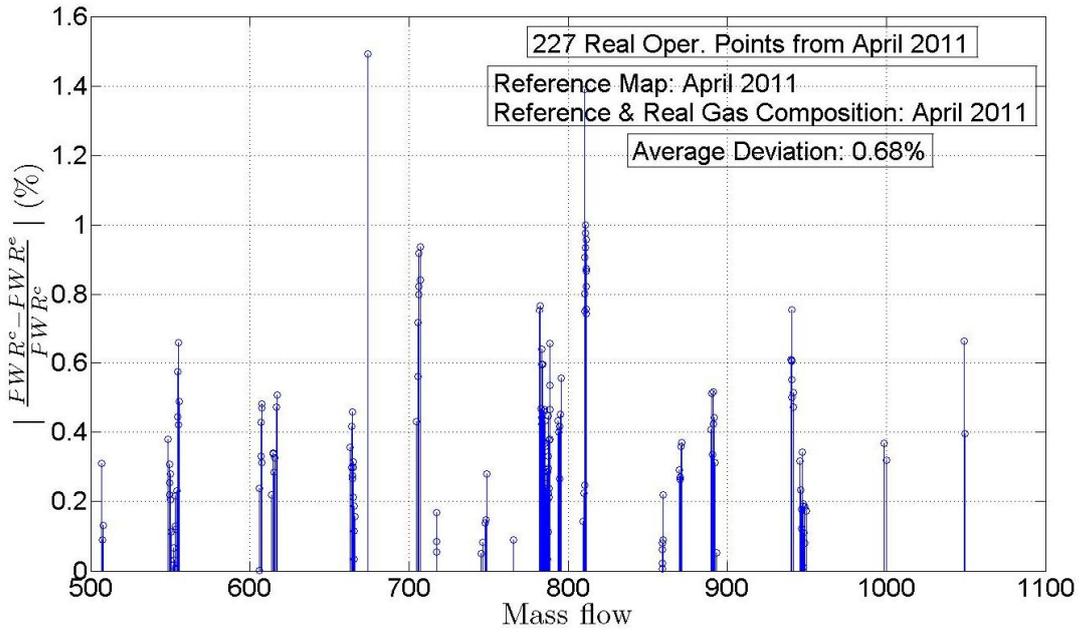

Figure 6 Corrected Gas Power for Operating Data from 2011.

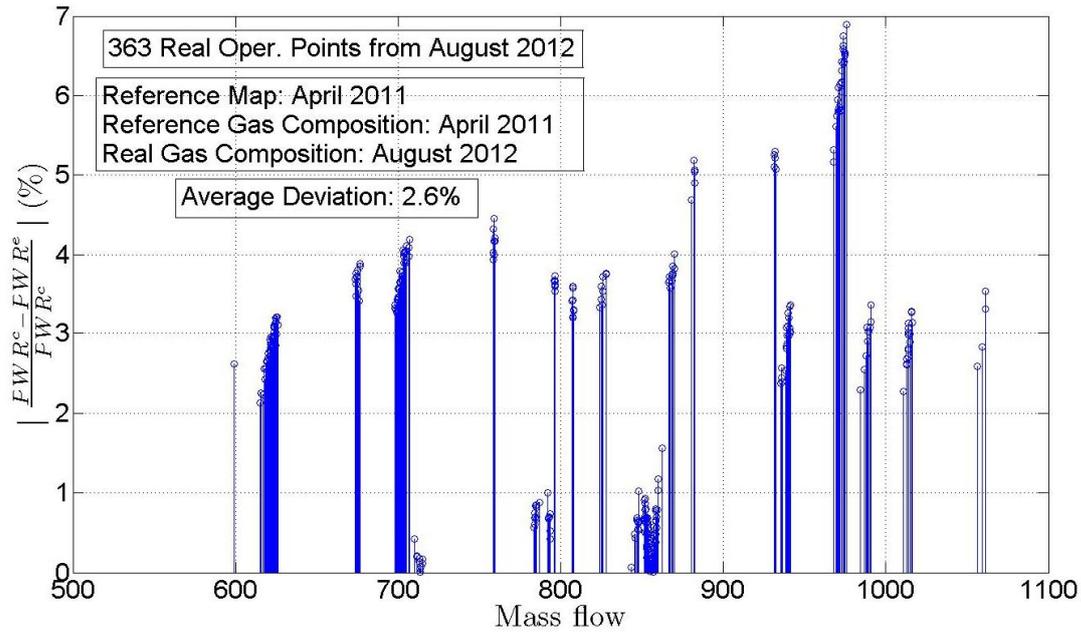

Figure 7 Corrected Gas Power for Operating Data from 2012.

## 6. CONCLUSIONS

In this paper, we have proposed an iterative method, which enables to correct the actual compressor performance under the real operating conditions to the equivalent performance valid for the given reference conditions. The general procedures and detailed theory of the corrections have been described. The accuracy of the method has been demonstrated by comparing the deviations between the corrected performance data and the expected values obtained from a reference map.

The corrections have been performed on three year measurements from real industry gas processing pump. The obtained average deviations of the corrected polytropic head from the expected one are: 3.67%, 0.97%, and 3.5% for the years 2009, 2011, and 2012. The obtained average deviations for the corrected gas power for these three years are: 2.08%, 0.68%, 2.6%.

It can be concluded that the correction method is practical and performs well over real industry data. The method returns very small deviations when simultaneously correcting the polytropic head, the gas power, and the polytropic efficiency.

The proposed correction procedure and theory enable to compare the actual compressor performance from different operating period fairly. In addition, it also allows comparing real performance with the reference values provided by compressor manufactures. With the helps from the corrections, the abnormal compressor performance and potential problems can be identified at an early stage so that failures or mandatory shutdown can be avoided, which therefore significantly reduces maintenance costs.

## APPENDIX

A. Calculations of the Polytropic Exponent and Specific Heat Ratio for Real-Gas Conditions

This appendix explains how to calculate the polytropic exponent and the specific heat ratio.

The first step is to calculate the thermodynamic properties from the given operating conditions, i.e., the given pressure $P$, temperature $T$, and gas composition. The compressibility factor $z$ and the heat capacity $c_p$ are the necessary thermodynamic properties needed for the purpose of this appendix and can be calculated according to [8]. The second step is to compute the required partial derivatives. The mathematic details of these partial derivatives can be found in [8, 11].

As can be seen in [7] the polytropic exponent for real-gas conditions is given by

$$n = \frac{1+X}{Y \cdot \left[ \frac{1}{k} \cdot \left( \frac{1}{\eta} + X \right) - \left( \frac{1}{\eta} - 1 \right) \right]}, \tag{A.1}$$

Where

$$X = \frac{T}{z} \cdot \frac{\partial z}{\partial T}, \tag{A.2}$$

$$Y = 1 - \frac{P}{z} \cdot \frac{\partial z}{\partial P}, \tag{A.3}$$

represent the two additional functions. These two functions are used to supplement the compressibility factor calculated under real-gas conditions. The expressions for the partial derivatives $\partial z / \partial T$ and $\partial z / \partial P$ in (A.2) and (A.3) can be found in [11].

The exponent

$$k = 1 - \frac{c_p}{c_p - T \cdot \frac{\partial P}{\partial T} \cdot \frac{\partial v}{\partial T}} \tag{A.4}$$

denotes the specific heat ratio, and $c_p$ is known as the heat capacity of hydrocarbon mixtures. The expressions for the partial derivatives $\partial P / \partial T$ and $\partial v / \partial T$ can be found in [8].

For convenience, we denote the calculations of these exponents by the function 'ExponentCalc', i.e., $(X, Y, k, n) = ExponentCalc(P, T, c_p, z, \eta, \text{Gas composition})$.

It has to be mentioned that the expressions presented in this appendix can be used to calculate the polytropic exponent and specific heat ratio either at inlet or discharge, one just need to substituting the relevant required parameters from inlet or discharge.

B. Prof of Equations (11) and (12)

Based on the deductions in [p.180, 6], the ratio of the volumetric flow at inlet to the flow at discharge, denoted by $r_v$, can be written as

$$r_v = \left( \frac{P_2}{P_1} \right)^{\frac{1}{n}}. \tag{B.1}$$

Similarly, the ratio of the corrected volumetric flow at inlet to the corrected flow at discharge is given by

$$r_v^c = \left(\frac{P_2^c}{P_1^c}\right)^{\frac{1}{n^c}}. \tag{B.2}$$

The second assumption made in Section 2 indicates that the ratio of the volumetric flow ratio at inlet and discharge remains the same before and after corrections, i.e.,

$$r_v = r_v^c. \tag{B.3}$$

Then, the corrected discharge pressure can be derived from equations (B.1) and (B.2) and expressed as

$$P_2^c = P_1^c \cdot \left(\frac{P_2}{P_1}\right)^{\frac{n^c}{n}}, \tag{B.4}$$

Starting from the relation between volumetric flow and temperature presented in [p.180, 6], equation (12) can be obtained.